\documentclass[10pt,conference]{IEEEtran}
\IEEEoverridecommandlockouts
\usepackage{graphicx}
\usepackage{graphics}
\usepackage{epsfig}
\usepackage{amsfonts}
\usepackage{amssymb}
\usepackage{amscd}
\usepackage{array}
\usepackage{eqparbox}
\usepackage[cmex10]{amsmath}
\newtheorem{theorem}{Theorem}[section]

\newcommand{\qed}{\nobreak \ifvmode \relax \else
      \ifdim\lastskip<1.5em \hskip-\lastskip
      \hskip1.5em plus0em minus0.5em \fi \nobreak
      \vrule height0.75em width0.5em depth0.25em\fi}
\def\BibTeX{{\rm B\kern-.05em{\sc i\kern-.025em b}\kern-.08em
    T\kern-.1667em\lower.7ex\hbox{E}\kern-.125emX}}

\begin{document}
\title{Two Classes of  Broadcast Channels With Side-Information: Capacity Outer Bounds}
\author{K. G. Nagananda$^1$, Chandra R Murthy$^2$ and Shalinee Kishore$^1$
\\\\
\begin{tabular}[h]{ccc}
  $^1$Dept. of ECE & $^2$Dept. of ECE    \\
  Lehigh University & Indian Institute of Science    \\
  Bethlehem, PA 18105, USA & Bangalore 560012, India \\
  \{kgn209, skishore\}@lehigh.edu & cmurthy@ece.iisc.ernet.in
\end{tabular}
}

\maketitle
\begin{abstract}
In this paper, we derive outer bounds on the capacity region of two classes of the general two-user discrete memoryless broadcast channels with side-information at the transmitter. The first class comprises the classical broadcast channel where a sender transmits two independent messages to two receivers. A constraint that each message must be kept confidential from the unintended receiver constitutes the second class. For both classes, the conditional distribution characterizing the channel depends on a state process and the encoder has side-information provided to it in a noncausal manner. For the first class of channels, an outer bound is derived employing techniques used to prove the converse theorem for the Gel'fand-Pinsker's channel with random parameters; the bounds are tight for individual rate constraints, but can be improved upon for the sum rate. The technique for deriving outer bounds for the second class of channels hinges on the confidentiality requirements; we also derive a genie-aided outer bound, where a hypothetical genie gives the unintended message to a receiver which treats it as side-information during equivocation computation. For both classes of channels, Csisz\'{a}r's sum identity plays a central role in establishing the capacity outer bounds.
\end{abstract}

\section{Introduction}\label{sec:introduction}
An information-theoretic study of broadcast channels (BC) was first initiated by Cover in \cite{refcoverbroadcast1}. In the classical setting, the BC comprises a sender who wishes to transmit $k$ independent messages to $k$ noncooperative receivers. The largest known inner bound on the capacity region when $k=2$ has been derived by Marton \cite{refmartonbroadcast1}, while outer bounds for this scenario have appeared in \cite{refnairbcout1}, \cite{refliangbcout1}. Several variants of this classical setting have also received considerable attention. One of the most prominent variants is BC with side-information, where the conditional probability distribution characterizing the channel depends on a state process, and where the channel side-information is available at the transmitter or at the receiver or at both ends. Capacity inner bounds for the two-user BC with noncausal side-information at the transmitter have been derived by Steinberg and Shamai in \cite{refsteinberg1}, where Marton's achievability scheme has been extended to state dependent channels. In \cite{refsteinberg2}, inner and outer bounds are derived for the degraded BC with noncausal side-information at the transmitter; the capacity region is derived when side-information is provided to the encoder in a causal manner. The capacity region for BC with receiver side-information is derived in \cite{refkramerbroadcast1}, where a \emph{genie} provides each receiver with the message that it need not decode.

In a wireless network paradigm, the issue of information security has attracted concerns mainly due to the broadcast nature of the wireless medium. In the information theory literature, capacity results/bounds have been derived for both point-to-point and several multiuser networks with security/confidentiality constraints. For BC with confidential messages, Csisz\'{a}r and K\"{o}rner derived capacity bounds for the two-user scenario \cite{refcsiszar1}, where the sender transmits a private message to $\mathrm{receiver}$ $1$ and a common message to both receivers, while keeping the private message confidential from $\mathrm{receiver}$ $2$. Capacity bounds have been derived in \cite{refliu1} for broadcasting two independent messages to two receivers, by keeping each message confidential from the unintended receiver. See \cite{refliang2} for an exhaustive coverage of papers in this area of research.

\subsection{Our contribution}\label{subsec:ourcontribution}
In this paper, we derive outer bounds on the capacity region of the two-user discrete memoryless BC with
\begin{enumerate}
\item noncausal side-information at the transmitter. This constitutes BC of $\mathrm{Class~I}$. An inner bound for this class of BC was derived in \cite{refsteinberg1}.
\item noncausal side-information at the transmitter and confidentiality constraints, where two independent messages are transmitted to two receivers such that each message must be kept confidential from the unintended receiver. This constitutes BC of $\mathrm{Class~II}$. An inner bound for this setting has been derived by the authors in \cite{refnandaicc2012}.
\end{enumerate}
For channels of $\mathrm{Class~I}$, we employ techniques used to prove the converse theorem for the Gel'fand-Pinsker's channel with random parameters \cite{refgelfand}, while for channels of $\mathrm{Class~II}$, confidentiality constraints are utilized to derive outer bounds. A genie-aided outer bound is also derived for channels of $\mathrm{Class~II}$. For both classes of channels, Csisz\'{a}r's sum identity \cite{refcsiszarbook} plays a central role in establishing the capacity outer bounds. The remainder of the paper is organized as follows. In Section \ref{sec:systemmodel}, we introduce the notation used and provide a mathematical model for the discrete memoryless version of the channels considered in this paper. In Section \ref{sec:mainresults}, we describe an outer bound to the capacity region of the two classes of channels and suggest a tighter outer bound for the sum rate of $\mathrm{Class~II}$ channels. We conclude the paper in Section \ref{sec:conclusions}. The proofs of the theorems are relegated to appendices.

\section{Channel Model \& Notation}\label{sec:systemmodel}
We denote the channel model by $\mathrm{C}_i; i=1,2$ is the model index. Calligraphic letters are used to denote finite sets, with a probability function defined on them. Uppercase letters denote random variables (RV), while boldface uppercase letters denote a sequence of RVs. Lowercase letters are used to denote particular realizations of RVs, and boldface lowercase letters denote $\mathrm{N}-$length vectors. $\mathrm{N}$ is the number of channel uses and $n=1,\dots,\mathrm{N}$ denotes the channel index. The sender is denoted $\mathrm{S}$ and the receivers are denoted $\mathrm{D}_t$; $t=1,2$ is the receiver index. Discrete random variables (RV) defined on finite sets $X \in \mathcal{X}$ and $Y_t \in \mathcal{Y}_t$ denote the channel input and outputs, respectively. The encoder of $\mathrm{S}$ is supplied with side-information $\mathbf{w} \in \mathcal{W}^\mathrm{N}$, in a noncausal manner. The channel is assumed to be memoryless and is characterized by the conditional distribution $p(\mathbf{y}_1,\mathbf{y}_2|\mathbf{x},\mathbf{w}) = \prod_{n=1}^{\mathrm{N}}p(y_{1,n},y_{2,n}|x_{n},w_{n})$. During proofs of outer bounds, the following notation will be useful for sequences of RVs: Consider $\mathbf{Y}_1 \triangleq (Y_{1,1}, \dots, Y_{1,\mathrm{N}})$. Then, $\mathbf{Y}_{1}^{n-1} \triangleq (Y_{1,1}, \dots, Y_{1,n-1})$ and $\mathbf{Y}^{\mathrm{N}}_{1,n+1} \triangleq (Y_{1,n+1}, \dots, Y_{1,\mathrm{N}})$.

To transmit its messages, $\mathrm{S}$ generates two RVs $M_{t} \in \mathcal{M}_{t}$, where $\mathcal{M}_{t} = \{1,\dots,2^{\mathrm{N}R_{t}}\}$ denotes a set of message indices. Without loss of generality, $2^{\mathrm{N}R_{t}}$ is assumed to be an integer, with $R_{t}$ being the transmission rate intended to $\mathrm{D}_t$. $M_{t}$ denotes the message, $\mathrm{S}$ intends to transmit to $\mathrm{D}_t$, and is assumed to be independently generated and uniformly distributed over the finite set $\mathcal{M}_{t}$. Integer $m_{t}$ is a particular realization of $M_{t}$ and denotes the message-index. 

For the channel $\mathrm{C}_1$, a $((2^{\mathrm{N}R_{1}},2^{\mathrm{N}R_{2}}),\mathrm{N},P_e^{(\mathrm{N})})$ code comprises:
\begin{enumerate}
\item $\mathrm{N}$ encoding functions $f$, such that $\mathbf{x} = \mathbf{f}(m_1,m_2,\mathbf{w})$,
\item Two decoders - $g_t: \mathcal{Y}^\mathrm{N}_t \rightarrow \mathcal{M}_{t}$.
\end{enumerate}
For the channel $\mathrm{C}_2$, a $((2^{\mathrm{N}R_{1}},2^{\mathrm{N}R_{2}}),\mathrm{N},P_e^{(\mathrm{N})})$ code comprises:
\begin{enumerate}
\item A stochastic encoder, which is defined by the matrix of conditional probabilities $\phi(\mathbf{x}|m_{1},m_{2},\mathbf{w})$, such that $\sum_{\mathbf{x}}\phi(\mathbf{x}|m_{1},m_{2},\mathbf{w}) = 1$. Here, $\phi(\mathbf{x}|m_{1},m_{2},\mathbf{w})$ denotes the probability that a pair of message-indices $(m_{1},m_{2})$ is encoded as $\mathbf{x} \in \mathcal{X}^\mathrm{N}$ to be transmitted by $\mathrm{S}$, in the presence of noncausal side-information $\mathbf{w}$.
\item Two decoders - $g_t: \mathcal{Y}^\mathrm{N}_t \rightarrow \mathcal{M}_{t}$.
\end{enumerate}

The average probability of decoding error for the code, averaged over all codes, is
$P_{\mathrm{C}_i,e}^{(\mathrm{N})} = \max \{P_{e,1}^{(\mathrm{N})},P_{e,2}^{(\mathrm{N})}\}$, with,
\begin{eqnarray*}
P_{e,t}^{(\mathrm{N})} = \sum_{\mathbf{m}}\sum_{\mathbf{w}\in \mathcal{W}^\mathrm{N}}\frac{1}{2^{\mathrm{N}[R_{1}+R_{2}]}}\text{Pr}\left[g_t(\mathcal{Y}_t^\mathrm{N}) \neq m_{t}|\mathbf{m},\mathbf{w}~ \text{sent}\right],
\end{eqnarray*}
where $\mathbf{m} = (m_{1},m_{2})$. A rate pair $(R_{1},R_{2})$ is said to be achievable for the channel $\mathrm{C}_i$, if there exists a sequence of $((2^{\mathrm{N}R_{1}},2^{\mathrm{N}R_{2}}),\mathrm{N},P_{\mathrm{C}_i,e}^{(\mathrm{N})})$ codes, $\forall \delta > 0$ and sufficiently small, such that $P_{\mathrm{C}_i,e}^{(\mathrm{N})} \leq \delta$ as $\mathrm{N} \rightarrow \infty$. Since the channel $\mathrm{C}_2$ has confidentiality requirements, $(R_1,R_2)$ must also satisfy the following weak-secrecy constraints \cite{refmaurer2} to be considered achievable for the channel $\mathrm{C}_2$:
\begin{eqnarray}
\mathrm{N}R_{1} - H(M_{1}|Y_2) \leq \mathrm{N}\delta, \label{eq:security1}   \\
\mathrm{N}R_{2} - H(M_{2}|Y_1) \leq \mathrm{N}\delta, \label{eq:security2}
\end{eqnarray}
where $H(\mathrm{x}|\mathrm{y})$ is the conditional entropy of $\mathrm{x}$ given $\mathrm{y}$. For both channels, the capacity region is defined as the closure of the set of all achievable rate pairs $(R_{1},R_{2})$.

\section{Main Results}\label{sec:mainresults}
\subsection{$\mathrm{Class~I}$: Broadcast channels with side-information}\label{subsec:bcsideinfo}
For the channel $\mathrm{C}_1$, we consider the set $\mathcal{P}_1$ of all joint probability distributions $p_1(w,v_1,v_2,x,y_1,y_2)$ that is constrained to factor as follows:
\begin{eqnarray*}
p_1(w,v_1,v_2,x,y_1,y_2) = p(w)p(v_1,v_2|w)\\ \times p(x|w,v_1,v_2)p(y_1,y_2|x).
\end{eqnarray*}
For a given $p_1(.)\in \mathcal{P}_1$, an outer bound for $\mathrm{C}_1$ is described by the set $\mathcal{R}_{1,\text{out}}(p_1)$, which is defined as the union of all rate pairs $(R_{1},R_{2})$ that simultaneously satisfy $(\ref{eq:C1outboundR1})$ - $(\ref{eq:C1outboundR1plusR2})$.
\begin{eqnarray}
R_1 &\leq& I(V_1;Y_1) - I(W;V_1), \label{eq:C1outboundR1}\\
R_2 &\leq& I(V_2;Y_2) - I(W;V_2), \label{eq:C1outboundR2}\\
\nonumber R_1 + R_2 &\leq& I(V_1;Y_1) + I(V_2;Y_2)\\ &&- I(W;V_1) - I(W;V_2). \label{eq:C1outboundR1plusR2}
\end{eqnarray}
\begin{theorem}\label{thm:conversethmC1}
Let $\mathcal{C}_1$ denote the capacity region of the channel $\mathrm{C}_1$.  Let $\mathcal{R}_{1,\text{out}} = \bigcup_{p_1(.)\in \mathcal{P}_1}\mathcal{R}_{1,\text{out}}(p_1)$. The region $\mathcal{R}_{1,\text{out}}$ is an outer bound for $\mathrm{C}_1$, i.e., $ \mathcal{C}_1 \subseteq \mathcal{R}_{1,\text{out}}$.
\end{theorem}
The proof of Theorem \ref{thm:conversethmC1} can be found in Appendix \ref{appendix:proofC1}.

\subsection{$\mathrm{Class~II}$: Broadcast channels with side-information \& confidential messages}\label{subsec:securebcsideinfo}
For the channel $\mathrm{C}_2$, we consider the set $\mathcal{P}_2$ of all joint probability distributions $p_2(w,u,v_1,v_2,x,y_1,y_2)$ that is constrained to factor as follows:
\begin{eqnarray*}
p_2(w,u,v_1,v_2,x,y_1,y_2) = p(w)p(u)p(v_1,v_2|w,u)\\ \times p(x|w,v_1,v_2)p(y_1,y_2|x).
\end{eqnarray*}
For a given $p_2(.)\in \mathcal{P}_2$, an outer bound for $\mathrm{C}_2$ is described by the set $\mathcal{R}_{2,\text{out}}(p_2)$, which is defined as the union over all distributions $p_2(.)$ of all rate pairs $(R_{1},R_{2})$ that simultaneously satisfy $(\ref{eq:minC2outboundR1})$ - $(\ref{eq:minC2outboundR1plusR2})$.
\begin{eqnarray}
R_1 &\leq& \min[I_1, I^{\ast}_1], \label{eq:minC2outboundR1}\\
R_2 &\leq& \min[I_2, I^{\ast}_2], \label{eq:minC2outboundR2}\\
R_1 + R_2 &\leq& \min[I_{12}, I^{\ast}_{12}], \label{eq:minC2outboundR1plusR2}
\end{eqnarray}
where $I_1,\dots,I^{\ast}_{12}$ are given by $(\ref{eq:C2outboundR1})$ - $(\ref{eq:C2outboundR1plusR2genie})$, respectively. Note that, $I^{\ast}_1$, $I^{\ast}_2$ and $I^{\ast}_{12}$ are genie-aided outer bounds, where a genie gives $\mathrm{D}_1$ message $M_2$, while $\mathrm{D}_2$ computes the equivocation using $M_2$ as side-information. The auxiliary RVs $U$, $V_1$ and $V_2$ are constrained to satisfy the following Markov chains: $U\rightarrow V_1 \rightarrow X$ and $U\rightarrow V_2 \rightarrow X$.
\begin{figure*}[ht!]
\centering
\begin{eqnarray}
R_1 &\leq& I(V_1;Y_1|U) - I(V_1;Y_2|U) + H(W|U,V_1)=I_1, \label{eq:C2outboundR1}\\
R_2 &\leq& I(V_2;Y_2|U) - I(V_2;Y_1|U) + H(W|U,V_2)=I_2, \label{eq:C2outboundR2}\\
R_1 + R_2 &\leq& I(V_1;Y_1|U) + I(V_2;Y_2|U) - I(V_1;Y_2|U) - I(V_2;Y_1|U) + H(W|U,V_1) + H(W|U,V_2)=I_{12}.\label{eq:C2outboundR1plusR2}\\
\nonumber \\
R_1 &\leq& I(V_1;Y_1|U,V_2) - I(V_1;Y_2|U,V_2) + H(W|U,V_1,V_2) = I^{\ast}_1, \label{eq:C2outboundR1genie}\\
R_2 &\leq& I(V_2;Y_2|U,V_1) - I(V_2;Y_1|U,V_1) + H(W|U,V_1,V_2) = I^{\ast}_2, \label{eq:C2outboundR2genie}\\
R_1 + R_2 &\leq& I(V_1;Y_1|U,V_2) + I(V_2;Y_2|U,V_1) - I(V_1;Y_2|U,V_2) - I(V_2;Y_1|U,V_1) + 2H(W|U,V_1,V_2) = I^{\ast}_{12}.\label{eq:C2outboundR1plusR2genie}
\end{eqnarray}
\hrulefill
\end{figure*}
\begin{theorem}\label{thm:conversethmC2}
Let $\mathcal{C}_2$ denote the capacity region of the channel $\mathrm{C}_2$.  Let $\mathcal{R}_{2,\text{out}} = \bigcup_{p_2(.)\in \mathcal{P}_2}\mathcal{R}_{2,\text{out}}(p_2)$. The region $\mathcal{R}_{2,\text{out}}$ is an outer bound for $\mathrm{C}_2$, i.e., $ \mathcal{C}_2 \subseteq \mathcal{R}_{2,\text{out}}$.
\end{theorem}
The proof of Theorem \ref{thm:conversethmC2} can be found in Appendices \ref{appendix:proofC2} and \ref{appendix:proofC2genie}.

\subsection{Inner bounds for $\mathrm{Class~I}$ \& $\mathrm{Class~II}$ channels}
For a given $p_1(.)\in \mathcal{P}_1$, a lower bound on the capacity region for $\mathrm{C}_1$ is described by the set $\mathcal{R}_{1,\text{in}}(p_1)$, which is defined as the union over all distributions $p_1(.)$ of the convex-hull of the set of all rate pairs $(R_{1},R_{2})$ that simultaneously satisfy $(\ref{eq:rateregionBCsideinfoR1})$ - $(\ref{eq:rateregionBCsideinfoR1plusR2})$. It was first characterized by Steinberg and Shamai \cite[Theorem 1]{refsteinberg1}.
\begin{eqnarray}
R_1 &\leq& I(V_1;Y_1) - I(W;V_1),\label{eq:rateregionBCsideinfoR1}\\
R_2 &\leq& I(V_2;Y_2) - I(W;V_2),\label{eq:rateregionBCsideinfoR2}\\
\nonumber R_1 + R_2 &\leq& I(V_1;Y_1) + I(V_2;Y_2)\\ && - I(V_1;V_2) - I(V_1,V_2;W). \label{eq:rateregionBCsideinfoR1plusR2}
\end{eqnarray}
Comparing $(\ref{eq:C1outboundR1})$ - $(\ref{eq:C1outboundR1plusR2})$ with $(\ref{eq:rateregionBCsideinfoR1})$ - $(\ref{eq:rateregionBCsideinfoR1plusR2})$, we see that the outer bounds are tight for the individual rate constraints, $R_1$ and $R_2$. However, the bound on $R_1+R_2$ can be improved upon.

For a given $p_2(.)\in \mathcal{P}_2$, an inner bound on the capacity region for $\mathrm{C}_2$ is described by the set $\mathcal{R}_{2,\text{in}}(p_2)$, which is defined as the union over all distributions $p_2(.)$ of the convex-hull of the set of all rate pairs $(R_{1},R_{2})$ that simultaneously satisfy $(\ref{eq:rateregionR1})$ - $(\ref{eq:rateregionR1plusR2})$. For proof, see \cite[Theorem 3.1]{refnandaicc2012}.
\begin{eqnarray}
\nonumber R_1 &\leq& I(V_1;Y_1|U)\\ &&- \max [I(V_1;Y_2|U,V_2),I(W;V_1|U)],\label{eq:rateregionR1}\\
\nonumber R_2 &\leq& I(V_2;Y_2|U)\\ &&- \max [ I(V_2;Y_1|U,V_1),I(W;V_2|U)],\label{eq:rateregionR2}\\
\nonumber R_1 + R_2 &\leq& I(V_1;Y_1|U) + I(V_2;Y_2|U)\\ \nonumber  &&- I(V_1;Y_2|U,V_2) - I(V_2;Y_1|U,V_1)\\
 &&- I(V_1;V_2|U) - I(V_1,V_2;W|U). \label{eq:rateregionR1plusR2}
\end{eqnarray}

\subsection{A tighter bound on $R_1+R_2$ for $\mathrm{Class~II}$ channels}
For the channel $\mathrm{C}_2$, the outer bound on $R_1+R_2$ can be made tighter by following a simple procedure. From $(\ref{eq:C2outboundR1})$ - $(\ref{eq:C2outboundR1plusR2})$, we see that $R_1+R_2 \leq I_1 + I_2$, and from $(\ref{eq:C2outboundR1genie})$ - $(\ref{eq:C2outboundR1plusR2genie})$ we have $R_1+R_2 \leq I^{\ast}_1 + I^{\ast}_2$. Therefore,
\begin{eqnarray}
R_1+R_2 \leq \min [I_1 + I^{\ast}_2, I_2 + I^{\ast}_1]. \label{eq:sumratebound}
\end{eqnarray}
We show now that the bound $(\ref{eq:sumratebound})$ is tighter than $(\ref{eq:C2outboundR1plusR2})$ and $(\ref{eq:C2outboundR1plusR2genie})$. It is easy to see that
\begin{eqnarray}
\nonumber I_1 + I_2 = I^{\ast}_1 + I^{\ast}_2 + I(W;V_1|U,V_2) + I(W;V_2|U,V_1).
\end{eqnarray}
Consider
\begin{eqnarray}
\nonumber 2(I_1 + I_2) = 2[I^{\ast}_1 + I^{\ast}_2 + I(W;V_1|U,V_2) + I(W;V_2|U,V_1)],
\end{eqnarray}
which implies the following:
\begin{eqnarray}
\nonumber \min[I_1 + I^{\ast}_2, I_2 + I^{\ast}_1] &\leq& I_1 + I_2,\\
\nonumber \min[I_1 + I^{\ast}_2, I_2 + I^{\ast}_1] &\leq& I^{\ast}_1 + I^{\ast}_2.
\end{eqnarray}
Therefore, the sum rate bound given by $(\ref{eq:sumratebound})$ is tighter than $(\ref{eq:C2outboundR1plusR2})$ and $(\ref{eq:C2outboundR1plusR2genie})$.

\section{Conclusions}\label{sec:conclusions}
We derived capacity outer bounds for two classes of broadcast channels. $\mathrm{Class~I}$ channels comprised two-user BC with side-information provided to the transmitter in a noncausal manner. For this class of channels, an outer bound is derived employing techniques used to derive the converse theorem for Gel'fand-Pinsker's channels with random parameters. We showed that the bounds are tight for individual rate constraints, but the bound on the sum rate can be improved upon. BC with noncausal side-information at the transmitter and confidentiality constraints, where each message is kept confidential from the unintended receiver, constituted channels of $\mathrm{Class~II}$. For this class of channels, we derived two types of outer bounds; the genie-aided outer bound is derived by letting a hypothetical genie give the unintended message to a receiver, while that receiver computes equivocation treating the unintended message as side-information.

\appendices
\section{}\label{appendix:proofC1}
Here, we prove Theorem \ref{thm:conversethmC1}. $\forall \epsilon >0$ and sufficiently small; and for large $\mathrm{N}$, $R_1$ can be bounded as follows:
\begin{eqnarray}
\nonumber \mathrm{N}R_1 &=& H(M_1) = I(M_1;\mathbf{Y}_1^\mathrm{N})+H(M_1|\mathbf{Y}_1^\mathrm{N})\\
\nonumber &\stackrel{(a)}\leq& I(M_1;\mathbf{Y}_1^\mathrm{N}) + \mathrm{N}\epsilon \\
\nonumber &\stackrel{(b)}=& \sum^{\mathrm{N}}_{n=1}[H(Y_{1,n}|\mathbf{Y}^{n-1}_1) - H(Y_{1,n}|\mathbf{Y}^{n-1}_1,M_1)] + \mathrm{N}\epsilon \\
\nonumber &\stackrel{(c)}\leq& \sum^{\mathrm{N}}_{n=1}[H(Y_{1,n}) - H(Y_{1,n}|\mathbf{Y}^{n-1}_1,M_1)] + \mathrm{N}\epsilon \\
\nonumber &=& \sum^{\mathrm{N}}_{n=1}I(M_1,\mathbf{Y}^{n-1}_1;Y_{1,n}) + \mathrm{N}\epsilon \\
\nonumber &=& \sum^{\mathrm{N}}_{n=1}[I(M_1,\mathbf{Y}^{n-1}_1,\mathbf{W}^{\mathrm{N}}_{n+1};Y_{1,n})\\
\nonumber &&- I(\mathbf{W}^{\mathrm{N}}_{n+1};Y_{1,n}|M_1,\mathbf{Y}^{n-1}_1)] + \mathrm{N}\epsilon \\
\nonumber &\stackrel{(d)}=& \sum^{\mathrm{N}}_{n=1}[I(M_1,\mathbf{Y}^{n-1}_1,\mathbf{W}^{\mathrm{N}}_{n+1};Y_{1,n})\\
\nonumber &&- I(\mathbf{Y}^{n-1}_1;W_n|M_1,\mathbf{W}^{\mathrm{N}}_{n+1})] + \mathrm{N}\epsilon \\
\nonumber &\stackrel{(e)}=& \sum^{\mathrm{N}}_{n=1}[I(M_1,\mathbf{Y}^{n-1}_1,\mathbf{W}^{\mathrm{N}}_{n+1};Y_{1,n})\\
\nonumber &&- I(M_1,\mathbf{W}^{\mathrm{N}}_{n+1},\mathbf{Y}^{n-1}_1;W_n)] + \mathrm{N}\epsilon,
\end{eqnarray}
where $(a)$ follows from Fano's inequality \cite{refcoverbook}, $(b)$ follows from chain rule, $(c)$ follows from the fact that conditioning reduces entropy, $(d)$ follows from Csisz\'{a}r's sum identity and $(e)$ is due to the fact that $(M_1,\mathbf{W}^{\mathrm{N}}_{n+1})$ is independent of $W_n$.
Letting $V_{1,n}=(M_1,\mathbf{W}^{\mathrm{N}}_{n+1},\mathbf{Y}^{n-1}_1)$, we get
\begin{eqnarray}
\mathrm{N}R_1 \leq \sum^{\mathrm{N}}_{n=1}I(V_{1,n};Y_{1,n}) - I(V_{1,n};W_n) + \mathrm{N}\epsilon. \label{eq:appendproofC1R1}
\end{eqnarray}
Proceeding in a similar manner and letting $V_{2,n}=(M_2,\mathbf{W}^{\mathrm{N}}_{n+1},\mathbf{Y}^{n-1}_2)$, we get the following bound on $R_2$:
\begin{eqnarray}
\mathrm{N}R_2 \leq \sum^{\mathrm{N}}_{n=1}I(V_{2,n};Y_{2,n}) - I(V_{2,n};W_n) + \mathrm{N}\epsilon. \label{eq:appendproofC1R2}
\end{eqnarray}
A bound on the sum rate $R_1+R_2$ is obtained by using the fact that $M_1$ and $M_2$ are independent and following the same procedure used to bound $R_1$ and $R_2$, to get
\begin{eqnarray}
\nonumber \mathrm{N}(R_1+R_2) \leq \sum^{\mathrm{N}}_{n=1}[I(V_{1,n};Y_{1,n}) + I(V_{2,n};Y_{2,n})\\
\nonumber   - I(V_{1,n};W_n) - I(V_{2,n};W_n)]\\ \!\!\!\!\! + 2\mathrm{N}\epsilon. \label{eq:appendproofC1R1plusR2}
\end{eqnarray}

\section{}\label{appendix:proofC2}
Here, we prove Theorem \ref{thm:conversethmC2}. $\forall \epsilon >0$ and sufficiently small; and for large $\mathrm{N}$, $R_1$ can be bounded as follows:
\begin{eqnarray}
\nonumber \mathrm{N}R_1 &=& H(M_1) = I(M_1;\mathbf{Y}_1^\mathrm{N})+H(M_1|\mathbf{Y}_1^\mathrm{N})\\
\nonumber &\stackrel{(a)}\leq& I(M_1;\mathbf{Y}_1^\mathrm{N}) + \mathrm{N}\epsilon \\
\nonumber &\stackrel{(b)}\leq& I(M_1;\mathbf{Y}_1^\mathrm{N}) - I(M_1;\mathbf{Y}_2^\mathrm{N}) + 2\mathrm{N}\epsilon \\
\nonumber &=& \sum^{\mathrm{N}}_{n=1}[I(M_1;Y_{1,n}|\mathbf{Y}^{\mathrm{N}}_{1,n+1}) - I(M_1;Y_{2,n}|\mathbf{Y}^{n-1}_2)]\\ \nonumber && + 2\mathrm{N}\epsilon \\
\nonumber &\stackrel{(c)}=& \sum^{\mathrm{N}}_{n=1}[I(M_1,\mathbf{Y}^{n-1}_2;Y_{1,n}|\mathbf{Y}^{\mathrm{N}}_{1,n+1})\\
\nonumber &&- I(M_1,\mathbf{Y}^{\mathrm{N}}_{1,n+1};Y_{2,n}|\mathbf{Y}^{n-1}_2)] + 2\mathrm{N}\epsilon \\
\nonumber &\stackrel{(d)}=& \sum^{\mathrm{N}}_{n=1}[I(M_1;Y_{1,n}|\mathbf{Y}^{\mathrm{N}}_{1,n+1},\mathbf{Y}^{n-1}_2)\\
\nonumber &&- I(M_1;Y_{2,n}|\mathbf{Y}^{\mathrm{N}}_{1,n+1},\mathbf{Y}^{n-1}_2)] + 2\mathrm{N}\epsilon \\
\nonumber &\leq& \sum^{\mathrm{N}}_{n=1}[I(M_1,W_n;Y_{1,n}|\mathbf{Y}^{\mathrm{N}}_{1,n+1},\mathbf{Y}^{n-1}_2)\\
\nonumber &&- I(M_1;Y_{2,n}|\mathbf{Y}^{\mathrm{N}}_{1,n+1},\mathbf{Y}^{n-1}_2)] + 2\mathrm{N}\epsilon \\
\nonumber &\stackrel{(e)}=& \sum^{\mathrm{N}}_{n=1}[I(M_1;Y_{1,n}|\mathbf{Y}^{\mathrm{N}}_{1,n+1},\mathbf{Y}^{n-1}_2)\\
\nonumber &&+ I(W_n;Y_{1,n}|M_1,\mathbf{Y}^{\mathrm{N}}_{1,n+1},\mathbf{Y}^{n-1}_2)\\
\nonumber &&-I(M_1;Y_{2,n}|\mathbf{Y}^{\mathrm{N}}_{1,n+1},\mathbf{Y}^{n-1}_2)] + 2\mathrm{N}\epsilon \\
\nonumber &=& \sum^{\mathrm{N}}_{n=1}[I(M_1;Y_{1,n}|\mathbf{Y}^{\mathrm{N}}_{1,n+1},\mathbf{Y}^{n-1}_2)\\
\nonumber &&+ H(W_n|M_1,\mathbf{Y}^{\mathrm{N}}_{1,n+1},\mathbf{Y}^{n-1}_2)\\
\nonumber &&- H(W_n|M_1,Y_{1,n},\mathbf{Y}^{\mathrm{N}}_{1,n+1},\mathbf{Y}^{n-1}_2)\\
\nonumber &&-I(M_1;Y_{2,n}|\mathbf{Y}^{\mathrm{N}}_{1,n+1},\mathbf{Y}^{n-1}_2)] + 2\mathrm{N}\epsilon \\
\nonumber &\leq& \sum^{\mathrm{N}}_{n=1}[I(M_1;Y_{1,n}|\mathbf{Y}^{\mathrm{N}}_{1,n+1},\mathbf{Y}^{n-1}_2)\\
\nonumber &&+ H(W_n|M_1,\mathbf{Y}^{\mathrm{N}}_{1,n+1},\mathbf{Y}^{n-1}_2)\\
\nonumber &&-I(M_1;Y_{2,n}|\mathbf{Y}^{\mathrm{N}}_{1,n+1},\mathbf{Y}^{n-1}_2)] + 2\mathrm{N}\epsilon,
\end{eqnarray}
where $(a)$ is from Fano's inequality, $(b)$ is from confidentiality constraints, $(c)$ and $(d)$ follow from Csisz\'{a}r's sum identity and $(e)$ is the chain rule for mutual information. Letting $U_n = (\mathbf{Y}^{\mathrm{N}}_{1,n+1},\mathbf{Y}^{n-1}_2)$ and $V_{1,1} = \dots = V_{1,\mathrm{N}} = M_1$ we get
\begin{eqnarray}
\nonumber \mathrm{N}R_1 &\leq& \sum^{\mathrm{N}}_{n=1}[I(V_{1,n};Y_{1,n}|U_n)+H(W_n|U_n,V_{1,n})\\
 &&- I(V_{1,n};Y_{2,n}|U_n)] + 2\mathrm{N}\epsilon. \label{eq:appendproofC2R1}
\end{eqnarray}
Proceeding in a similar fashion and letting $V_{2,1} = \dots = V_{2,\mathrm{N}} = M_2$, $R_2$ can be bounded as follows:
\begin{eqnarray}
\nonumber \mathrm{N}R_2 &\leq& \sum^{\mathrm{N}}_{n=1}[I(V_{2,n};Y_{2,n}|U_n)+H(W_n|U_n,V_{2,n})\\
 &&- I(V_{2,n};Y_{1,n}|U_n)] + 2\mathrm{N}\epsilon. \label{eq:appendproofC2R2}
\end{eqnarray}
A bound on the sum rate $R_1+R_2$ is obtained by using the fact that $M_1$ and $M_2$ are independent and following the same procedure used to bound $R_1$ and $R_2$.
\begin{eqnarray}
\nonumber \mathrm{N}(R_1+R_2) \leq \sum^{\mathrm{N}}_{n=1}[I(V_{1,n};Y_{1,n}|U_n) + I(V_{2,n};Y_{2,n}|U_n)\\
\nonumber - I(V_{1,n};Y_{2,n}|U_n)] - I(V_{2,n};Y_{1,n}|U_n)\\ + H(W_n|U_n,V_{1,n})
 + H(W_n|U_n,V_{2,n}) + 4\mathrm{N}\epsilon. \label{eq:appendproofC2R1plusR2}
\end{eqnarray}

\section{}\label{appendix:proofC2genie}
For the channel $\mathrm{C}_2$, consider a hypothetical genie which gives $\mathrm{D}_1$ message $M_2$, while $\mathrm{D}_2$ computes the equivocation using $M_2$ as side-information. $\forall \epsilon >0$ and sufficiently small; and for large $\mathrm{N}$, $R_1$ can be upper bounded as follows:
\begin{eqnarray}
\nonumber \mathrm{N}R_1 &=& H(M_1) \leq H(M_1|\mathbf{Y}_2^\mathrm{N}) + \mathrm{N}\epsilon\\
\nonumber &\leq& H(M_1,M_2|\mathbf{Y}_2^\mathrm{N}) + \mathrm{N}\epsilon\\
\nonumber &=& H(M_1|\mathbf{Y}_2^\mathrm{N},M_2) + H(M_2|\mathbf{Y}_2^\mathrm{N})  + \mathrm{N}\epsilon\\
\nonumber &\leq& H(M_1|\mathbf{Y}_2^\mathrm{N},M_2) + \mathrm{N}\epsilon\\
\nonumber &\leq& H(M_1|\mathbf{Y}_2^\mathrm{N},M_2) - H(M_1|\mathbf{Y}_1^\mathrm{N}) + \mathrm{N}\epsilon \\
\nonumber &\stackrel{(a)}\leq& H(M_1|\mathbf{Y}_2^\mathrm{N},M_2) - H(M_1|\mathbf{Y}_1^\mathrm{N},M_2) + \mathrm{N}\epsilon\\
\nonumber &\leq& I(M_1;\mathbf{Y}_1^\mathrm{N}|M_2) - I(M_1;\mathbf{Y}_2^\mathrm{N}|M_2) + 2\mathrm{N}\epsilon \\
\nonumber &=& \sum^{\mathrm{N}}_{n=1}[I(M_1;Y_{1,n}|\mathbf{Y}^{\mathrm{N}}_{1,n+1},M_2)\\ \nonumber &&- I(M_1;Y_{2,n}|\mathbf{Y}^{n-1}_2,M_2)] + 2\mathrm{N}\epsilon \\
\nonumber &\stackrel{(b)}=& \sum^{\mathrm{N}}_{n=1}[I(M_1,\mathbf{Y}^{n-1}_2;Y_{1,n}|\mathbf{Y}^{\mathrm{N}}_{1,n+1},M_2)\\
\nonumber &&- I(M_1,\mathbf{Y}^{\mathrm{N}}_{1,n+1};Y_{2,n}|\mathbf{Y}^{n-1}_2,M_2)] + 2\mathrm{N}\epsilon \\
\nonumber &\stackrel{(c)}=& \sum^{\mathrm{N}}_{n=1}[I(M_1;Y_{1,n}|\mathbf{Y}^{\mathrm{N}}_{1,n+1},\mathbf{Y}^{n-1}_2,M_2)\\
\nonumber &&- I(M_1;Y_{2,n}|\mathbf{Y}^{\mathrm{N}}_{1,n+1},\mathbf{Y}^{n-1}_2,M_2)] + 2\mathrm{N}\epsilon \\
\nonumber &\leq& \sum^{\mathrm{N}}_{n=1}[I(M_1,W_n;Y_{1,n}|\mathbf{Y}^{\mathrm{N}}_{1,n+1},\mathbf{Y}^{n-1}_2,M_2)\\
\nonumber &&- I(M_1;Y_{2,n}|\mathbf{Y}^{\mathrm{N}}_{1,n+1},\mathbf{Y}^{n-1}_2,M_2)] + 2\mathrm{N}\epsilon \\
\nonumber &=& \sum^{\mathrm{N}}_{n=1}[I(M_1;Y_{1,n}|\mathbf{Y}^{\mathrm{N}}_{1,n+1},\mathbf{Y}^{n-1}_2,M_2)\\
\nonumber &&+ I(W_n;Y_{1,n}|M_1,\mathbf{Y}^{\mathrm{N}}_{1,n+1},\mathbf{Y}^{n-1}_2,M_2)\\
\nonumber &&-I(M_1;Y_{2,n}|\mathbf{Y}^{\mathrm{N}}_{1,n+1},\mathbf{Y}^{n-1}_2,M_2)] + 2\mathrm{N}\epsilon \\
\nonumber &=& \sum^{\mathrm{N}}_{n=1}[I(M_1;Y_{1,n}|\mathbf{Y}^{\mathrm{N}}_{1,n+1},\mathbf{Y}^{n-1}_2,M_2)\\
\nonumber &&+ H(W_n|M_1,\mathbf{Y}^{\mathrm{N}}_{1,n+1},\mathbf{Y}^{n-1}_2,M_2)\\
\nonumber &&- H(W_n|M_1,Y_{1,n},\mathbf{Y}^{\mathrm{N}}_{1,n+1},\mathbf{Y}^{n-1}_2,M_2)\\
\nonumber &&-I(M_1;Y_{2,n}|\mathbf{Y}^{\mathrm{N}}_{1,n+1},\mathbf{Y}^{n-1}_2,M_2)] + 2\mathrm{N}\epsilon \\
\nonumber &\leq& \sum^{\mathrm{N}}_{n=1}[I(M_1;Y_{1,n}|\mathbf{Y}^{\mathrm{N}}_{1,n+1},\mathbf{Y}^{n-1}_2,M_2)\\
\nonumber &&+ H(W_n|M_1,\mathbf{Y}^{\mathrm{N}}_{1,n+1},\mathbf{Y}^{n-1}_2,M_2)\\
\nonumber &&-I(M_1;Y_{2,n}|\mathbf{Y}^{\mathrm{N}}_{1,n+1},\mathbf{Y}^{n-1}_2,M_2)] + 2\mathrm{N}\epsilon,
\end{eqnarray}
where $(a)$ follows since the genie gives $\mathrm{D}_1$ message $M_2$, $(b)$ and $(c)$ follow from Csisz\'{a}r's sum identity.
Letting $U_n = (\mathbf{Y}^{\mathrm{N}}_{1,n+1},\mathbf{Y}^{n-1}_2)$, $V_{1,1} = \dots = V_{1,\mathrm{N}} = M_1$ and $V_{2,1} = \dots = V_{2,\mathrm{N}} = M_2$, $R_1$ can be bounded as
\begin{eqnarray}
\nonumber \mathrm{N}R_1 \leq \sum^{\mathrm{N}}_{n=1}[I(V_{1,n};Y_{1,n}|U_n,V_{2,n})\\ \nonumber +H(W_n|U_n,V_{1,n},V_{2,n})
- I(V_{1,n};Y_{2,n}|U_n,V_{2,n})]\\ + 2\mathrm{N}\epsilon. \label{eq:appendproofC2R1genie}
\end{eqnarray}
Similarly,
\begin{eqnarray}
\nonumber \mathrm{N}R_2 \leq \sum^{\mathrm{N}}_{n=1}[I(V_{2,n};Y_{2,n}|U_n,V_{1,n})\\ \nonumber +H(W_n|U_n,V_{1,n},V_{2,n})
- I(V_{2,n};Y_{1,n}|U_n,V_{1,n})] \\ + 2\mathrm{N}\epsilon. \label{eq:appendproofC2R2genie}
\end{eqnarray}
To bound the sum rate, we use the fact that $M_1$ and $M_2$ are independent to get
\begin{eqnarray}
\nonumber \mathrm{N}(R_1+R_2) \leq \sum^{\mathrm{N}}_{n=1}[I(V_{1,n};Y_{1,n}|U_n,V_{2,n})\\ + I(V_{2,n};Y_{2,n}|U_n,V_{1,n})
\nonumber  - I(V_{1,n};Y_{2,n}|U_n,V_{2,n})\\ \nonumber  - I(V_{2,n};Y_{1,n}|U_n,V_{1,n})
+ 2H(W_n|U_n,V_{1,n},V_{2,n})\\ + 4\mathrm{N}\epsilon. \label{eq:appendproofC2R1plusR2genie}
\end{eqnarray}

Finally, a time sharing RV $Q$, which is uniformly distributed over $\mathrm{N}$ symbols and independent of all the RVs is introduced for the single letter characterization of the above derived outer bounds. Applying the procedure similar to the one presented in \cite[Chapter 15.3.4]{refcoverbook} on $(\ref{eq:appendproofC1R1})$ - $(\ref{eq:appendproofC1R1plusR2})$, $(\ref{eq:appendproofC2R1})$ - $(\ref{eq:appendproofC2R1plusR2})$ and $(\ref{eq:appendproofC2R1genie})$ - $(\ref{eq:appendproofC2R1plusR2genie})$, we get the outer bounds $(\ref{eq:C1outboundR1})$ - $(\ref{eq:C1outboundR1plusR2})$ and $(\ref{eq:minC2outboundR1})$ - $(\ref{eq:minC2outboundR1plusR2})$.

\bibliographystyle{IEEEtran}
\bibliography{IEEEabrv,icc2012}

\end{document}